\documentclass[12pt]{iopart}
\usepackage{subfig}
\usepackage{graphicx}

\begin{document}

\title[]{Macroscopic modeling of strain-rate dependent energy dissipation of
	superelastic SMAs considering destabilization of martensitic lattice \footnote{Accepted version of the paper published in Smart Materials and Structures, 29(2):025005, 2020, DOI: 10.1088/1361-665X/ab5e42}}
\author{A. Kaup, O. Altay and S. Klinkel}

\address{Department of Civil Engineering, RWTH Aachen University, \\
	 Mies-van-der-Rohe-Str. 1, 52074 Aachen, Germany}
\eads{\mailto{altay@lbb.rwth-aachen.de}}
\vspace{12pt}
\begin{indented}
	\item[]July 2019
\end{indented}

\begin{abstract}
Superelastic shape-memory alloys (SMAs) are unique smart materials with a considerable energy dissipation potential for dynamic loadings with varying strain-rates. The energy dissipation arises from a hysteretic phase transformation of the polycrystalline atomic grid structure. In fact, the nucleation from austenite to martensite phase and vice versa exhibits a strong thermomechanical coupling. The hysteresis depends on the latent heat generated by the austenitic-martensitic transformation and the convection of that heat. Due to the thermomechanical coupling, the martensitic nucleation stress level and thus the propagation of martensitic phase fronts strongly depends on the material temperature. High strain-rate interferes with the release of the latent heat to the environment and determines quantity, position and propagation of martensitic transformation bands. Lastly, high strain-rate reduces the hysteresis surface. The propagation velocity and quantity of martensitic bands have an impact on martensitic phase stability. The degree of atomic disorder and accordingly the change in entropy influences the reverse phase transformation from martensite to austenite. In other words, the stability of the martensitic state affects the stress level of the reverse transformation. However, in phenomenological superelastic SMA models, which are used to simulate energy dissipation behavior, the effects of the rate-dependent entropy change are not considered in particular.

To incorporate the rate-dependent entropy change, we improved a one-dimensional numerical model by introducing an additional control variable in the free energy formulation for solid-solid phase transformation of superelastic SMAs. In this model, the observed effects of the strain-rate on the reverse transformation are taken into account by calculating the rate-dependent entropy change. A comparison of the numerical results with the experimental data shows that the model calculates the dynamic superelastic hysteresis of SMAs more accurately.
\end{abstract}

\section{Introduction}

Polycrystalline shape-memory alloys (SMAs) are applied in various fields in engineering. SMAs with initial superelasticity are already extensively used in aeronautic, automotive and healthcare engineering. Furthermore, interest in SMAs as a damper system is growing in civil engineering \cite{Fang}. In general, SMA-based engineering applications experience predominantly strain-rate variated loadings. Especially in civil engineering both seismic events and wind cause dynamic loading on structures. In order to counteract this structural vibrations, several structural control strategies have been developed. Commonly used damping systems, such as steel-hysteresis-dampers, dissipate structural vibration energy by inelastic deformations. However, inelastic damping devices cannot recover deformations and thus suffer loss of damping effects. This problem motivates research into SMA-damping devices. In fact, superelastic SMA-damping devices can reduce structural vibrations significantly without irreversible deformations due to unique hysteretic behavior. 
The superelastic hysteretic material behavior of SMA is thermomechanically triggered and includes both martensitic transformation and reverse transformation. This study focuses on NiTi (Nickel titanium) SMA wires, which are the most commonly used SMAs in damping systems. For tensioned SMA wires, the stress level for phase transformation depends on strain-rate and loading history. The loading history has a decisive impact on the hysteretic behavior of NiTi wires if they are not trained. Accordingly, the in damping systems used NiTi wires are trained and thus the loading history does not decrease the superelastic hysteresis. Therefore, the loading history effects are not considered anymore in this study. However, the strain-rate strongly influences the thermomechanically triggered phase transformation. On the macroscopic level the nucleation from austenite to martensite is composed by the formation of martensitic transformation bands.  

The strain-rate dependent evolution of macroscopic transformation bands is essential to consider strain-rate effects in constitutive modeling of superelastic SMAs. One of the first investigations on the macroscopic formulation of phase transformation fronts in NiTi was conducted by Shaw \cite{Shaw}. Another seminal study regarding both the macroscopic and microscopic behavior of NiTi under cyclic loading was performed by Otsuka \cite{Otsuka}. Based on these investigations, Feng \cite{Feng} used electromechanical-opting testing machines and high resolution optical recorders to capture images of the material surface during varying strain-rate loadings. Pieczyska \cite{Pieczyska}, furthermore, used differential scanning calorimetry (DSC) to record the thermal evolution of the macroscopic phase transition bands, while additionally using infrared cameras to obtain thermograms. Another method to investigate the stress induced martensitic transformation is to use an acoustic emission sensor. In detail, Pieczyska \cite{Piecyska2} used an acoustic emission technique to measure the propagation of stress waves, which are related to rapid local volume changes. Precisely, in superelastic SMAs the evolution of martensitic transformation bands triggers a change in the material volume and thus produces stress waves. By comparing the acoustic emission to material temperature changes recorded by infrared camera and mechanical stress-strain diagrams it is possible to investigate the stress induced nucleation and propagation of martensitic phase fronts. Amini \cite{Amini,Amini2}, conducted nanoindentation tests on NiTi SMAs to investigate loading-rate dependent martensitic nucleation and phase front velocity. Therefore, he used a nanonindenter to apply cyclic loading, a 3D surface profiler to measure the surface roughness and DSC. Moreover, Xiao \cite{Xiao} investigated the strain field and the temperature field of NiTi specimen with 3D strain field cameras and infrared cameras. Also, Sun \cite{Sun} used both speed and thermal cameras to investigate the evolution and propagation of transformation domains for varying stretching rates. 

The strain-rate dependent constitutive modeling of superelastic SMAs for civil engineering applications is predominantly formulated as macroscopic material models. Brinson \cite{Brinson}, Graesser \cite{Graesser} and Auricchio \cite{Auricchio96}, among others, investigated fundamental macroscopic models. In the last decades, researchers developed several thermomechanical constitutive models for superelastic SMAs. In particular, Morin et. al. \cite{Morin} developed a three-dimensional constitutive model specialized for thermomechanical coupling. A three dimensional finite element model focused on the thermomechanical coupling in SMAs was also developed by Christ \cite{Christ}. Moreover, numerically simpler and thus computationally more efficient models were developed. Ren \cite{Ren} implemented a simple one-dimensional macroscopic model, which considers, based on the Greasser model, the strain-rate effects on cyclic loaded SMAs using the least square method. Furthermore, Zhu \cite{Zhu} developed a one-dimensional rate-dependent model for tensile stress, using a Runge-Kutta integration scheme. Also, Auricchio \cite{Auricchio} developed a thermomechanical rate-dependent, one-dimensional model based on his primary research. To cover the cycle-dependent behavior of superelastic SMAs, Sameallah \cite{Sam} proposes a further one-dimensional thermomechanical coupled constitutive model. Cisse \cite{Cisse} proposes a review paper on modeling techniques for superelastic SMAs. In general, all these models mainly differ in the free energy formulation, the solution algorithm and the evolutionary equation. Besides, Yu's \cite{Yu} one-dimensional model uses wave equations to cover the influence of the strain-rate on the phase front velocity during martensitic transformation. Furthermore, Acar et al. developed a numerical model based on neural networks and fuzzy logic, that covers both the rate- and temperature dependent behavior of SMAs \cite{Ozbulut}.

Both the evolution and propagation of martensitic phase fronts and the general thermomechanical coupling strongly influence the thermal evolution and the hysteretic behavior of dynamically excited SMAs. However, to our knowledge, simple and thus numerically efficient one-dimensional constitutive models developed so far do not explicitly consider the strain-rate dependent martensitic phase stability. In particular, the stress induced martensitic phase stability depends on the phase band nucleation and the velocity of phase front propagation. The quantity of martensitic transformation fronts and the propagation velocity of these is influenced by the strain-rate. Simultaneously, martensitic phase stability changes with the rate-dependent evolution of transition fronts.

This paper proposes an improved one-dimensional numerical model based on Auricchio's constitutive model \cite{Auricchio}, that allows integrating changes in phase stability depending on the strain-rate. The chosen numerical model of Auricchio is suitable for NiTi-wires as it is one-dimensional, and thus needs comparatively small number of input parameters to calculate the hysteretic behavior of SMA dampers. Furthermore, all the fundamental equations are thermomechanically consistent. The improved model considers the destabilization of martensitic lattice with macroscopic modeling approaches. Fast evolution of numerous phase bands, triggered by high strain-rate loads, destabilizes the martensitic lattice. The proposed constitutive model includes the energetic stability of the austenite phase state during increasing inherent temperature. For this purpose, we introduce a free-energy formulation, which calculates the entropy change in a strain-rate depending manner. The numerical model is validated by uniaxial cyclic tensile tests applied on trained NiTi-SMA wires.

The paper is divided into four sections. Section 2 introduces the dynamic material behavior of SMA and the experimental setup, which is used for the validation of the numerical model. Subsequently, Section 3 presents the constitutive model for superelastic SMA. Finally, the validation of the numerical calculations with the experimental results is presented in Section 4. 

\section{PRELIMINARY}
\subsection{Dynamic material properties}
The superelastic behavior of NiTi is thermomechanically coupled. For superelastic NiTi, the austenitic grid structure is energetically more effective.  Thus, energy, such as by tensile stress, has to be applied to initiate a phase transition. Auricchio \cite{Auricchio} defines in detail, the stress-induced phase transformation as indicated by the critical martensitic transformation start and finish points $R_s^{AM}$ and $R_f^{AM}$ and inversely by $R_s^{MA}$ and $R_f^{MA}$. Superscript $^{AM}$ indicates a transformation from austenite to martensite and $^{MA}$ vice versa. To incorporate the thermomechanical coupling, the critical transition point $R$ depends on the stress level $\sigma$ and the thermal level $T$. Indeed, the thermal level depends on the entropy $\eta$, however, here entropy is not rate-dependent and thus only a constant thermodynamic scaling factor. \begin{eqnarray} R = \sigma-T(\eta) \end{eqnarray} 
SMA's thermal evolution depends on the latent heat generation during the exothermic process, when the atomic grid restructures from a austenitic cubic to a martensitic monoclinic grid. Indeed, both the latent heat generation and the convection of the generated heat depend on the transformation-rate and on the diameter of the SMA specimen. As a result, especially the critical transformation points $R_f^{AM}$ and $R_s^{MA}$ shift upwards for a higher material temperature and consequently the slope of the hysteretic transformation area increases. In general, for higher external temperatures the whole hysteresis shifts upwards, because the high temperature austenite phase becomes more stable and thus more stress is needed to initiate the phase transformation. This phenomenon occurs regardless of the strain-rate and is considered by the Clausius-Clapeyron coefficient \cite{Casciati}. However, the latent heat generation induces inherent temperature changes, which influence the nucleation of the martensitic transformation bands. Thus, both the external and the internal temperature have to be considered.  
\begin{figure}
		\begin{center}
		\includegraphics[]{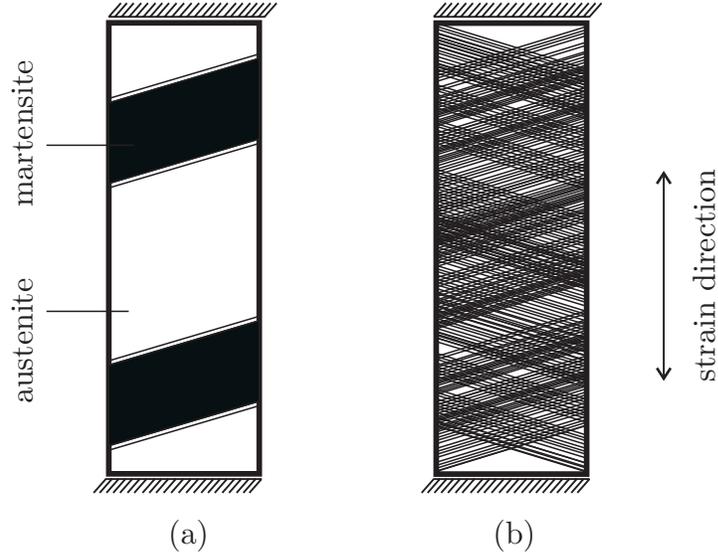}

	\caption{Schematic martensitic transformation induced band evolution for quasi-static (a) and dynamic (b) excitation. SMA wire fixed at both ends. Martensite-fractions are illustrated as black transformation bands.} 
	\label{band}
		\end{center}
\end{figure}
The phase transition induced thermal evolution is strain-rate and transformation-rate dependent and has an essential influence on the hysteresis surface. Additionally, the martensitic phase stability influences the reverse transformation. For instance, Fig.~\ref{band} illustrates the evolution of the martensitic transition fronts for both quasi-static (a) and dynamic excitations (b). In a quasi-static load-state, the growth of localized martensite bands determines the phase transition \cite{Xiao}. In general terms, for quasi-static excitation the stress field specifies the position of the transition fronts; hence, they are located next to the fixture points \cite{Shaw}. Evolution of the quasi-static martensite bands leads to wider bands and comparative phase-stable martensitic modes. In contrast, a higher strain-rate causes an increase of the nucleation sites in the material \cite{Amini}. An increase in the band quantity concurrently implies a decrease in the band width. Moreover, the front propagation velocity rises corresponding the loading-rate. The evolution of several coexistent transformation fronts destabilizes the intact structure of the material \cite{Shaw}. Consequently, the martensitic phase stability reduces for higher strain-rates, even if the amount of the martensite-volume in the specimen is comparable to the quasi-static case. In particular, the phase transition during high strain-rate loading induces an intrinsic instability of the transformation domains \cite{He}. The evolution of various both coexistent and existent transition fronts and the evolution of new defect zones in the material, such as inclusions or dislocations, are mutually dependent \cite{Amini}. In addition, the higher transformation-rate raises the degree of atomic disorder and the motion of the atoms. Indeed, a more unstable atomic state is concomitant with increasing entropy in the material \cite{Tatar}. As a consequence, the reverse transformation process is initiated early on a higher stress level, since the martensitic phase becomes too unstable and the material strives for the energetically more efficient and more stable austenite grid structure. It should be noted that the micromechanical processes during martensitic and reverse transformations are nonsymmetrical \cite{Piecyska2}. Summarized, high strain-rates cause a decrease of martensitic stability, which leads during unloading to a reverse transformation on the higher stress level.
\subsection{Experimental setup}
\begin{figure}[b]
	\begin{center}
		\includegraphics[]{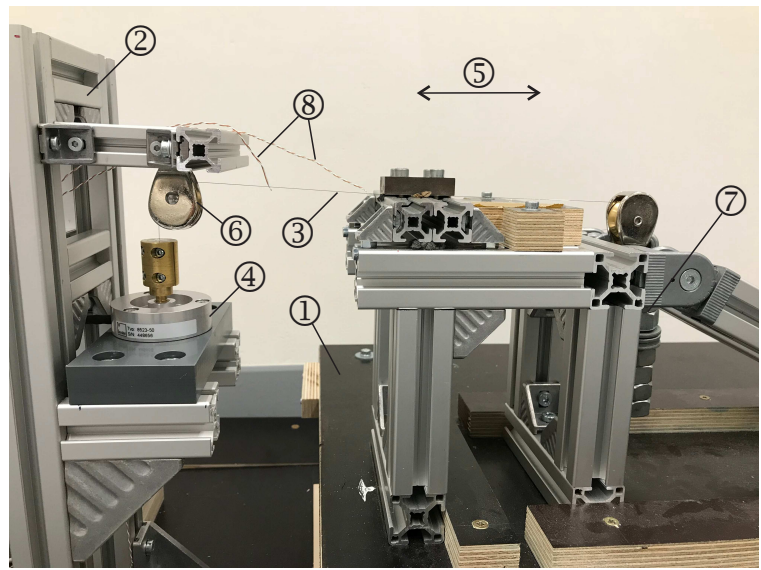}
		\caption{Utilized experimental setup, side-view. (1) shaking table, (2) test rig, (3) NiTi-wire, (4) load cell, (5), uniaxial strain direction, (6) pulley, (7) weights, (8) two T-type thermocouples.}
		\label{setup}
	\end{center}
\end{figure}
Fig.~\ref{setup} shows the utilized shaking table for the uniaxial cyclic-tensile tests. The uniaxial horizontal shaking table is 0.50 m x 0.50 m in size and can simulate both cyclic and stochastic loadings. The major components of the experimental setup are the shaking table (1) and a fixed test rig (2). NiTi wires (3) are attached to a vertically installed load cell (4). The distance from load cell to the right-side fixture is 150 mm and thus sets the NiTi wire length for the experimental investigations. The uniaxial direction movement (5) causes tensile stress on the NiTi wire. To avoid buckling, which is owing to pressure, a pulley (6) redirects the SMA-wire and ensure centric load application to the load cell. Additionally, weights (7) pre-stress the NiTi wire to induce martensitic transformation in a both faster and controlled manner. Moreover, two T-type thermocouples (8) are installed 5 and 60 mm distanced from the right-side fixture. This position enables local temperature evolution measurement and the identification of the local thermal differences. To control the table motion and to record the measurements a MATLAB/Simulink environment is used. A CAN-hub connects the MATLAB/Simulink code with the shaking table and introduces the excitation signals. Finally, all sensors are connected with BNC-wires via a National Instruments NI USB-6210 A/D converter to a PC.
\section{Rate-dependent constitutive SMA model}\label{sec:equ}

In this section, an advanced macroscopic model is introduced, which accounts for the entropy evolution to cover martensitic destabilization for high strain-rate tensile loads. The model is based on the subsequently summarized model of Auricchio \cite{Auricchio}.

To cover the thermomechanical behavior of superelastic SMAs, the constitutive model is based on thermodynamic potential state laws. To calculate the thermodynamic material behavior we use a set of internal and external variables referring to a homogenized volume element. In general, the internal variable $\xi$, representing the martensitic volume fraction in the material, is expressed for martensitic transformation in term of the following evolutionary equation
\begin{eqnarray}\label{eq:2}
\dot{\xi} = \beta^{AM}\xi \frac{\dot{F}}{\left(F-R_{f}^{AM}\right)^{2}}\;,
\end{eqnarray}
which can be controlled with the speed-parameter $\beta^{AM}$. This equation depends moreover on the driving force $F$ and the thermo-coupled limit stress level for martensitic transformation $R_{f}^{AM}$, where both depend on the external variables; the temperature $T$ and the uniaxial strain $\varepsilon$. To take into account the thermomechanical coupling, the constitutive model is based on a formulation of the free energy $\psi$. In particular, the derivation of $\psi$ originates from the thermo-elastic free energy formulation for single-phase materials of Raniecki and Bruhns \cite{Raniecki} and reads
\begin{eqnarray} \label{eq:3}
	\psi = \psi_{0}(T) + \psi^{e}\left(\varepsilon^{e}\right) - (T-T_{0})\eta^{e}\left(\varepsilon^{e},T\right) + C\left[\left(T-T_{0}\right) - T\ln \frac{T}{T_{0}}\right]\;.
\end{eqnarray} 
In this case, $\psi_{0}$ states the temperature-dependent free energy connection to the material's internal energy $u$ and entropy $\eta$ in the reference state and $\psi^{e}$ states the free energy dependency on elastic strain. Moreover, $C$ is the material heat capacity and $T_{0}$ the reference temperature. The heat capacity $C$ strongly influences the sensitivity of the material to thermal changes. Based on Eq.~\ref{eq:3}, Auricchio and Sacco proposed a free energy formulation \cite{Auricchio96} for solid-solid phase transformation in superelastic SMAs as,
\begin{eqnarray} \label{eq:4}
\psi = & \left[\left(u_{A}-T\eta_{A}\right) - \xi \left(\Delta u-T\Delta \eta\right)\right] +C\left[\left(T-T_{0}\right)-T\ln \frac{T}{T_{0}}\right] \nonumber \\
& +\frac{1}{2}E\varepsilon^{e^{2}} - \left(T-T_{0}\right)\varepsilon^{e}E\alpha\;,
\end{eqnarray}
where the subscript $-_{A}$ defines the austenite, $\Delta$ the constant difference between austenite and martensite values. Accordingly, $\Delta\eta$ is time-independent. Moreover, $\alpha$ is the thermal expansion and $E$ the elastic modulus, which follows the Reuss-scheme. The evolution of the elastic modulus depends on the martensitic volume ratio and thus ranges from the austenitic elastic modulus and the martensitic elastic modulus. Based on Eq.~\ref{eq:4}, the stress $\sigma$ and the temperature $T$ are defined by the corresponding derivations of the free energy formulation. Hence, $\sigma$ and $T$ are defined, based on the heat equation as
\begin{eqnarray}
\sigma = \frac{\partial\psi}{\partial\varepsilon^{e}} = E\varepsilon^{e}-E\alpha\left(T-T_{0}\right)
\end{eqnarray} 
and
\begin{eqnarray} \label{Eq T}
	C\dot{T} = T\frac{\partial^{2}\psi}{\partial T\partial\varepsilon^{e}}\dot{\varepsilon}+T\frac{\partial^{2}\psi}{\partial T\partial\xi}\dot{\xi}+\sigma\dot{\varepsilon} - \frac{\partial\psi}{\partial\varepsilon}\dot{\varepsilon}-\frac{\partial\psi}{\partial\xi}\dot{\xi}
	-\gamma \left(T-T_{ext}\right)\;.
\end{eqnarray}
To incorporate the rate-dependent entropy change, we rewrite $\psi_{0}(T)$, part of the free energy formulation of Eq.~\ref{eq:3}, by introducing the time-dependent entropy difference $\Delta\eta(t)$ to define $\psi$ as
\begin{eqnarray}
	\psi = & \left[\left(u_{A}-T\eta_{A}\right) - \xi \left(\Delta u-T\Delta \eta(t)\right)\right] + C\left[\left(T-T_{0}\right)-T\ln \frac{T}{T_{0}}\right] \nonumber\\
	&+\frac{1}{2}E\varepsilon^{e^{2}} - \left(T-T_{0}\right)\varepsilon^{e}E\alpha\;.
\end{eqnarray}
More in detail,  $\eta_{A}$ states for $t = 0$ the entropy level of the material in the pure austenite state. This state is also reached after each complete relief of strain and thus after each completed reverse transformation. Since we deal with thin wires, we neglect the thermal expansion factor $\alpha$ subsequently. Now, we introduce
\begin{eqnarray}\label{eta}
	\eta(t) = -\frac{\partial\psi}{\partial T} = \eta_{A} + \xi\Delta \eta(t) + C \ln \frac{T}{T_{0}}\;,
\end{eqnarray}
which updates $\eta$ not only depending on $\xi$ and $T$ but also on $\Delta\eta(t)$ with
\begin{eqnarray}\label{eta(t)}
\Delta\eta(t)=\eta(t)-\eta_A \;,
\end{eqnarray}
as an additional control variable, which defines the difference between the current entropy level $\eta(t)$ and the initial entropy level $\eta_{A}$ and thus considers a time-dependent and strain-rate dependent calculation of the entropy change. The rate-dependent entropy change calculation enables to consider the strain-rate effects on the reverse transformation. The increasing entropy is the measure to define the strain-rate dependent instability of the martensitic state. The rate-dependency of the calculation arises directly from $\xi$, which depends on the phase transformation-rate, and from $T$. In fact, the rate-dependency of $T$ comes directly from $\dot{\varepsilon}$ and $\dot{\xi}$ in Eq.~\ref{Eq T}. At this point, it is important to note, that with the existing temperature equation of the primary model the calculated evolution of the simulation temperature T is only qualitatively comparable to the experimentally measurable real temperature $T_{E}$. Experimental results show that the temperature increase is overestimated particularly for quasi-static excitations. In order to counteract this overestimation, we introduce the scaling factor $\kappa$, which reduces the temperature influence in the entropy calculation for the quasi-static load case. The differentiation between quasi-static and dynamic load cases depends on the entropy evolution. In detail, $\eta_{qs}$ is a threshold, which must be exceeded to calculate $\eta$ following Eq.~\ref{eta}. For quasi-static load cases, when $\eta < \eta_{qs}$, this paper proposes the entropy calculation dependent on $\kappa$ as follows,
\begin{eqnarray}\label{etaK} 
	\eta(t) = -\frac{\partial\psi}{\partial T} = \eta_{A} + \xi\Delta\eta(t) + \kappa C \ln \frac{T}{T_{0}}\;;\qquad for\;\eta < \eta_{qs}
\end{eqnarray}
Further research on the temperature equation is required to make the scaling factor $\kappa$ dispensable for future works. Nevertheless, the consideration of the rate-dependent entropy change enables already the inclusion of the influence of martensitic phase stability in the calculation. To consider the strain-rate effects, both on the initial and final reverse transition stress level and on the martensitic-austenitic hysteresis slope, we reformulate the thermo-coupled initial and final stress levels for the reverse transformation and the corresponding speed parameter as
\begin{eqnarray}
R_{s,f}^{MA} = \sigma_{s,f}^{MA}-T_{R}\frac{\eta(t)}{\varepsilon_{L}}\;,
\end{eqnarray}
and
\begin{eqnarray}\label{eq:11}
\beta^{MA} = \frac{\eta(t)}{\eta_{A}}\;.
\end{eqnarray}
The reformulation of the thermo-coupled reverse transformation stress levels comprises the strain-rate dependent entropy. Thus, the initial and final stress levels change by altering the entropy level depending on the strain-rate. The thermo-coupled initial stress level $R_{s}^{MA}$ determines the initial reverse transformation strain $\varepsilon_s^{MA}(\eta)$ and thus also the linear-elastic part of the austenitic-martensitic phase transition. More in detail, this material model is strain-driven and consequently we need strain values, which indicate the begin and finish of the reverse transformation. These strain levels are updated in every calculation step and thus change with varying entropy level to consider the martensitic instability for high strain-rates. Consideration of the entropy change for the initial and reverse strain levels of the austenitic transition leads to an increase of $\varepsilon_s^{MA}(\eta)$ and a decrease of $\varepsilon_f^{MA}(\eta)$ with increasing strain-rate. Next, the speed parameter $\beta^{MA}(\eta)$ increases with increasing strain-rate and thus the reverse transformation slope changes from a convex to a concave slope. In brief, the free energy reformulation enables a rate-dependent calculation of the entropy change and hence the introduction of both the entropy dependent speed parameter $\beta^{MA}(\eta)$ and the entropy dependent reverse transformation strains $\varepsilon_{s,f}^{MA}(\eta)$.  

\section{Results and discussion}
\begin{figure*}[]
	\centering
	\subfloat[]{
		\includegraphics[scale=0.6]{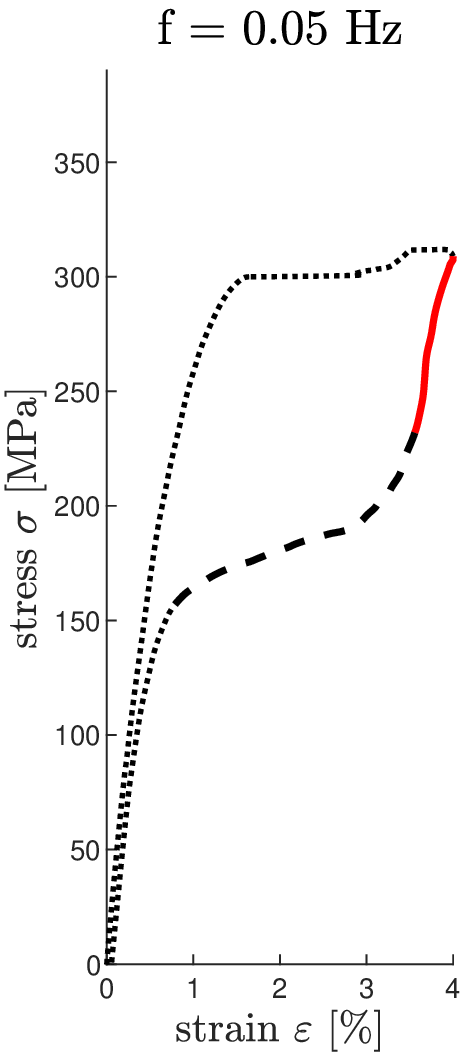}}
	\subfloat[]{%
		\includegraphics[scale=0.6]{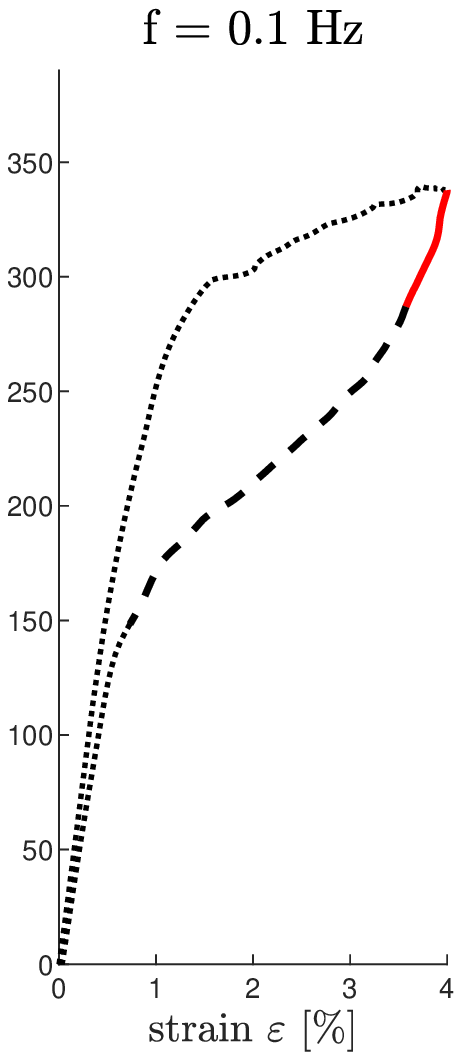}}
	\subfloat[]{%
		\includegraphics[scale=0.6]{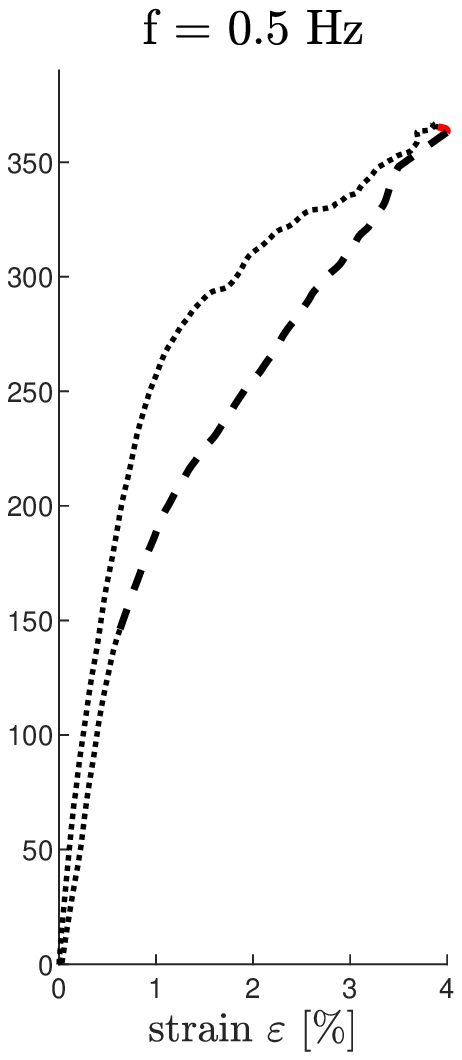}}
	\subfloat[]{%
		\includegraphics[scale=0.6]{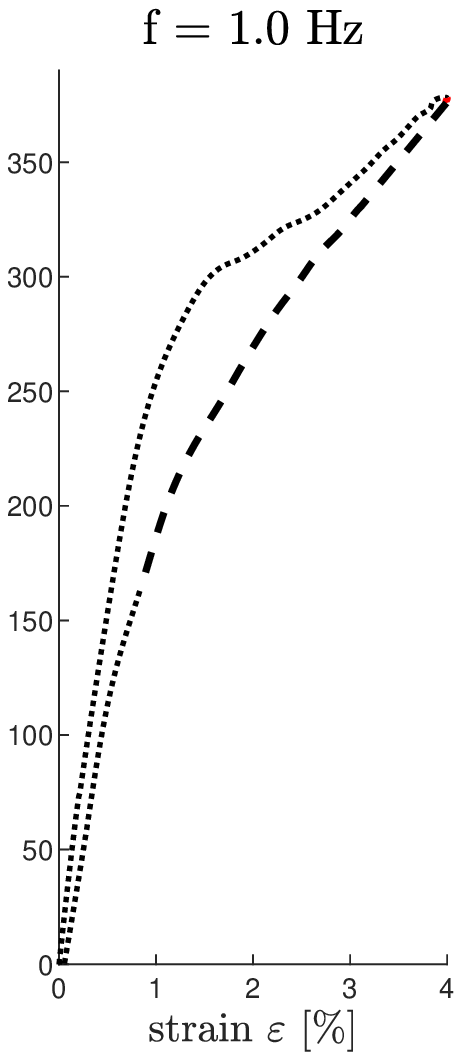}}
	\subfloat[]{%
		\includegraphics[scale=0.6]{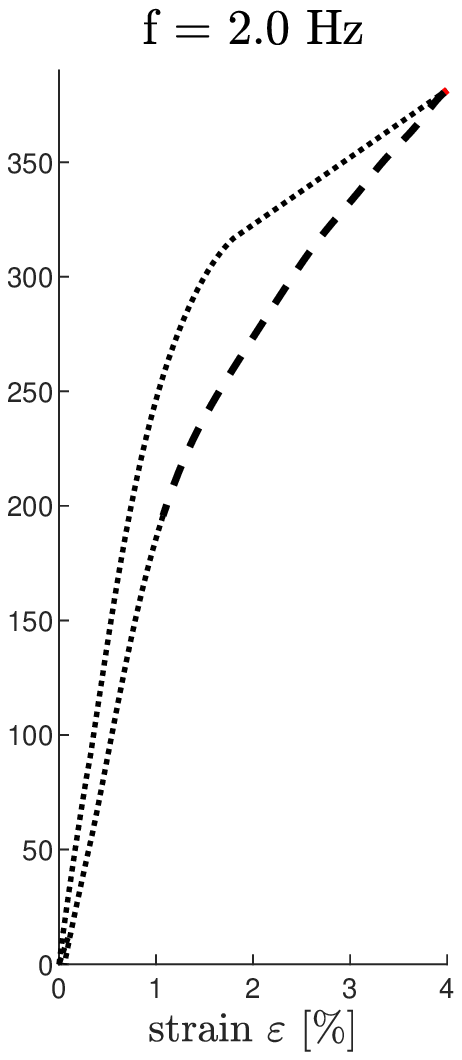}}
	\caption{Cyclic-tensile-tests on Ni-55.8\%-Ti-43.55\%-SMA wires. Red-marked solid line: martensitic linear-elastic section. Dashed line: reverse transformation. Strain amplitude: $\varepsilon=4\;\%$. Excitation frequency: $f=0.05 (a)/0.1 (b)/0.5 (c)/1.0 (d)/2.0 (e)$~Hz. Wire length: $L=150$~mm. Wire diameter: $d=0.2$~mm. Pre-stress: $\sigma_0=134.9$~MPa. Ambient temperature: $T=22.5\;{^\circ}$C.} 
	\label{Versuch} 	\centering
\end{figure*}

The following part will discuss both experimental and numerical results, by which the presented one-dimensional model will be validated. The experimental results in Fig.~\ref{Versuch} visualize the strain-rate effects during cyclic-tensile excitation with a strain amplitude of 4 \% on trained 0.2 mm diameter Ni-55.8\%-Ti-43.55\%-SMA-wires, produced by SAES Getters S.p.A. (SAES). In fact, we focus on the first loading cycle of each cyclic tensile test. Since we deal only with trained wires the hysteresis can be assumed not to change anymore for increasing amounts of load cycles itself. We choose still the first load cycle to be sure that every presented experimental result starts with the same initial temperature, without former self-heating due to cyclic loading. While comparing the whole hysteresis surface for frequencies of 0.05, 0.1, 0.5, 1.0 and 2.0 Hz, we visualize both the linear-elastic and the transformation section of the reverse transition. We observe an increase in the reverse phase transformation stress level and especially an upward shift in transformation plateau with increasing slope corresponding to the increasing strain-rate. As shown in previous studies, such as in \cite{ZhangX,ZhangY}, the significance of the effect changes depending on the thermodynamic properties of the SMA. In the results of Fig.~\ref{Versuch}, the chosen wire geometry and loading configuration magnifies the reverse transformation shift allowing us to show the entropy effect more clearly. The reverse transformation stress levels $\sigma_{s,f}^{MA}$ increase for increasing strain-rate. However, for high-strain rate tests, the exact position of the reverse transformation finish level cannot be observed clearly anymore, due to the direct phase transition of the investigated wires. Moreover, the decrease of the linear-elastic reverse transformation for increasing strain-rates becomes evident, as marked with a solid line in the plots. The decreasing linear-elastic section and the resulting reverse transformation on a higher stress level strongly influence the shape of the reverse transition hysteresis. The earlier the reverse transformation is initiated, the more directly the austenitic phase transition takes place. Concurrently, the fast and direct austenitic nucleation under high strain-rates causes a change of the reverse hysteresis slope. In particular, the slope changes from a convex Fig.~\ref{Versuch}(a) to a straight concave slope Fig.~\ref{Versuch}(e). This strain-rate dependent reverse transition shape is triggered by the decreasing martensitic phase-stability. The decreasing martensitic phase stability intensifies the strive to the energetically more effective austenite state. However, to our knowledge, the existing constitutive models do not consider these strain-rate dependent changes in the reverse transformation.

Subsequently, we present the effect of the entropy-change dependent calculation step by step, both of the reverse transformation strains $\varepsilon_{s,f}^{MA}(\eta)$ and the adjusted speed parameter $\beta^{MA}$. More precisely, the results are presented in the following order: no rate-dependent entropy change, entropy change adjustment of $\varepsilon_{s}^{MA}(\eta)$, entropy-change adjustment of $\varepsilon_{s,f}^{MA}(\eta)$ and $\beta^{MA}(\eta)$. 

The performed numerical calculations use a set of material and thermodynamic parameters. Because of its thermodynamic consistency, the material model can be used beside wires also for uniaxially stressed bars and cables by tuning the material and thermodynamic parameters of the model as listed in Table~\ref{Material}. We selected the material parameters, apart from the thermodynamic parameters, on the basis of quasi-static tensile tests taken both from the SAES data-sheet and self-conducted experimental results. The thermodynamic parameters were not determined experimentally, but chosen based on previous studies \cite{Auricchio}. Accordingly, we chose the following thermodynamic parameters appropriately to the behavior of the constitutive model in a physically consistent manner, see Table~\ref{Material}. The material parameter $\varepsilon_{L}$ is the maximum residual strain, $\gamma$ is a heat convection coefficient and $T_{ext}$ and $T_{R}$ are the external and the reference temperatures, respectively. The chosen thermodynamic parameters are in a similar range of values as in literature, such as in \cite{Auricchio}. $\Delta u$ and $\Delta \eta$ are used for the introduced entropy calculation and therefore differ from previous studies.
\begin{table}[]
	\caption{Material and thermodynamic parameters used for the constitutive model}
	\begin{indented}
	\item[]\begin{tabular}{@{}lll}
			\br
		    Parameter&Value  &  Unit \\
			\mr	
			$E_A$&32350  &MPa      \\
			$E_M$&18550  &MPa      \\
			$\varepsilon_{L}$&3.34  &\%      \\
			$\sigma_{s}^{AM}$&85  &MPa      \\
			$\sigma_{f}^{AM}$&305  &MPa      \\
			$\sigma_{s}^{MA}$&225  &MPa      \\
			$\sigma_{f}^{MA}$&200  &MPa      \\
			$\Delta u$& 1320   & MPa           \\ 
			$\eta_{A}$ & 0.0093 & MPaK$^{-1}$ \\ 
			$\eta_{qs}  $& 1.3 $\eta_A$ & MPaK$^{-1}$ \\ 
			$\kappa$ & 0.0035 & - \\ 
			$\gamma$& 0.1    & -                  \\ 
			C& 4.0    & MPaK$^{-1}$ \\ 
			$T_u$& 573    & K                  \\ 
			$T_{ext} = T_0 =T_R$& 295.65    & K                  \\ 
			$\alpha$& 0 & K$^{-1}$  \\ 
			\br	
			\end{tabular}
			\end{indented}
	\label{Material}
\end{table}

\begin{figure}[h]
	\centering
	\subfloat[]{
		\includegraphics[scale=0.63]{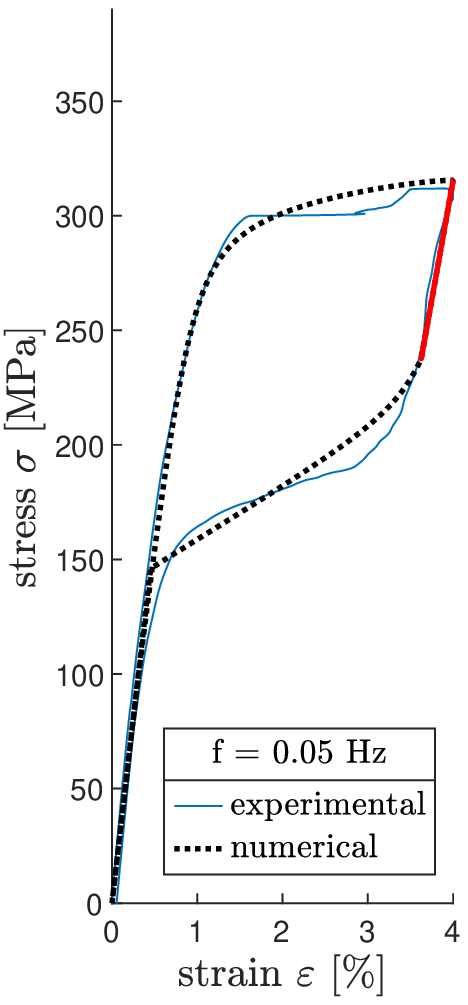}}
	\subfloat[]{%
		\includegraphics[scale=0.63]{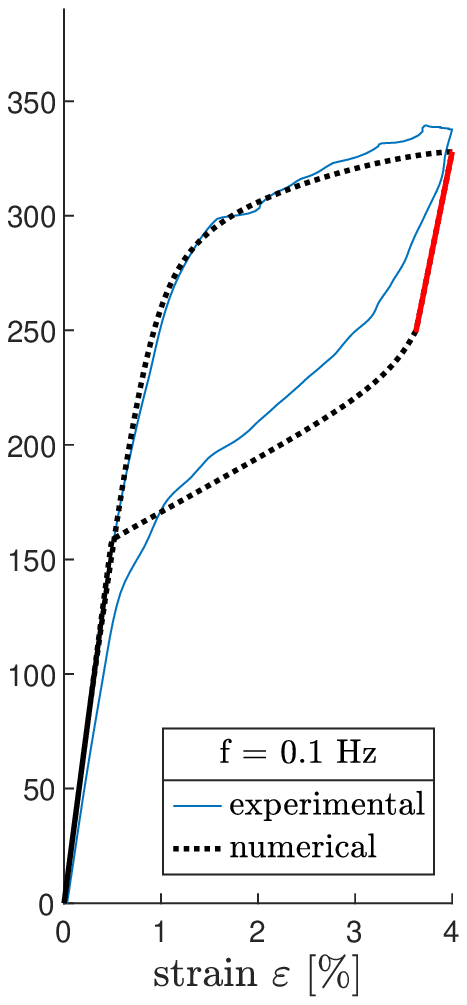}}
	\subfloat[]{%
		\includegraphics[scale=0.63]{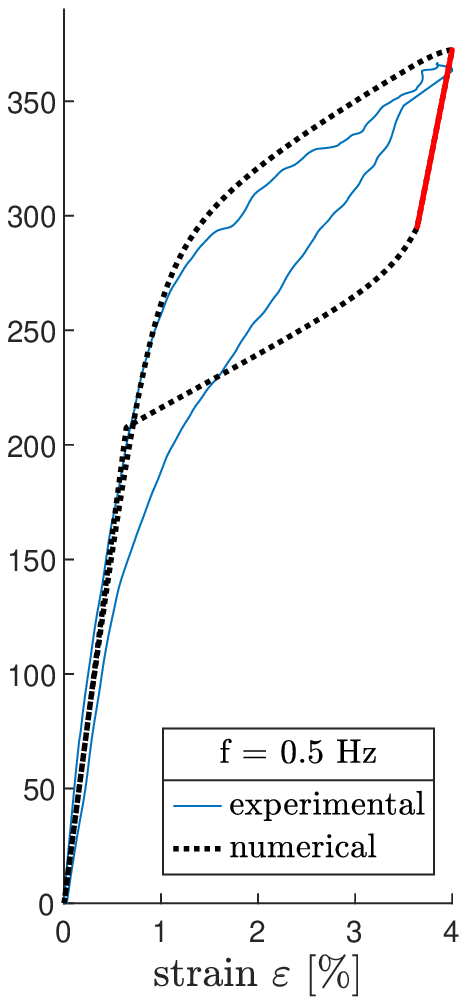}}
	\caption{Comparison of experimental data with numerical results excluding entropy-change effects. Red-marked solid line: martensitic linear-elastic section. $\varepsilon=4\;\%$. $f=0.05 (a)/0.1 (b)/0.5 (c)$~Hz. $L=150$~mm. $d=0.2$~mm. $\sigma_0=134.9$~MPa. $T=22.5\;{^\circ}$C.}
	\label{No}
\end{figure}

To show the influence of the rate-dependent entropy change on $\varepsilon_s^{MA}(\eta)$, we first compare the numerical results, excluding the entropy calculation, with the experimental results, see Fig.~\ref{No}. The linear-elastic part of the calculation is again separately marked. Indeed, the calculation of the martensitic-transformation is highly plausible for the three frequencies of 0.05, 0.1 and 0.5 Hz. Although the rate-dependent evolution of the simulation equals only qualitatively to the experimental temperature $T_{E}$, the model can represent the thermomechanical coupling for austenitic-martensitic transition. However, the linear-elastic reverse-transformation section does not change and the reverse transformation shape remains nearly the same. As a result, the hysteresis area for high strain-rate excitation is overestimated. Furthermore, the deviations in the calculation of the reverse transformation become evident.

\begin{figure}[h]
	\centering
	\subfloat[]{
		\includegraphics[scale=0.63]{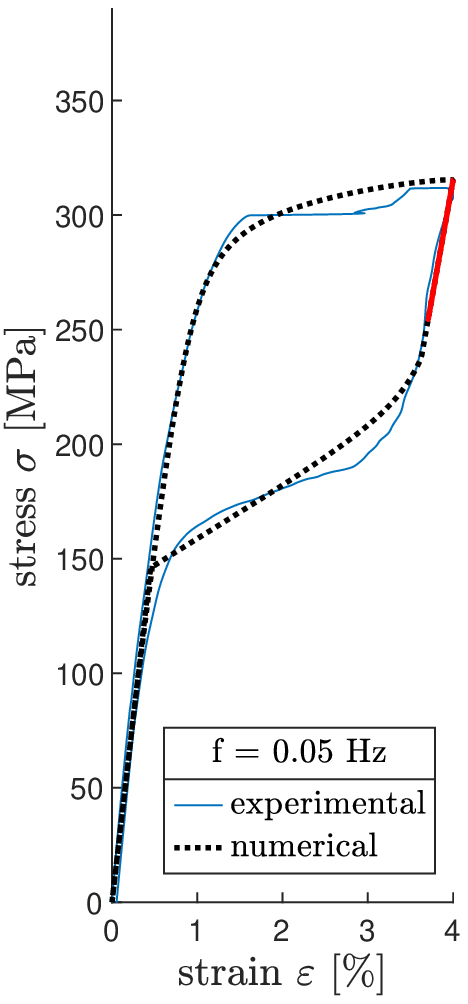}}
	\subfloat[]{%
		\includegraphics[scale=0.63]{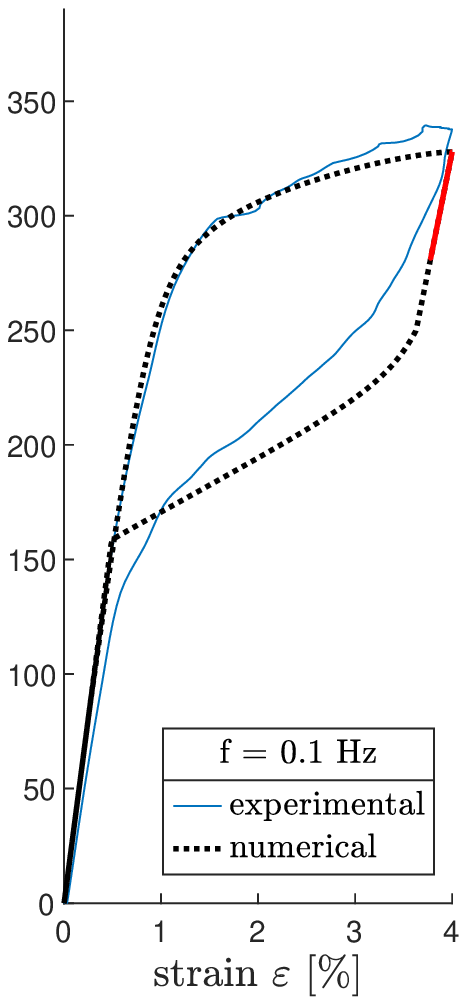}}
	\subfloat[]{%
		\includegraphics[scale=0.63]{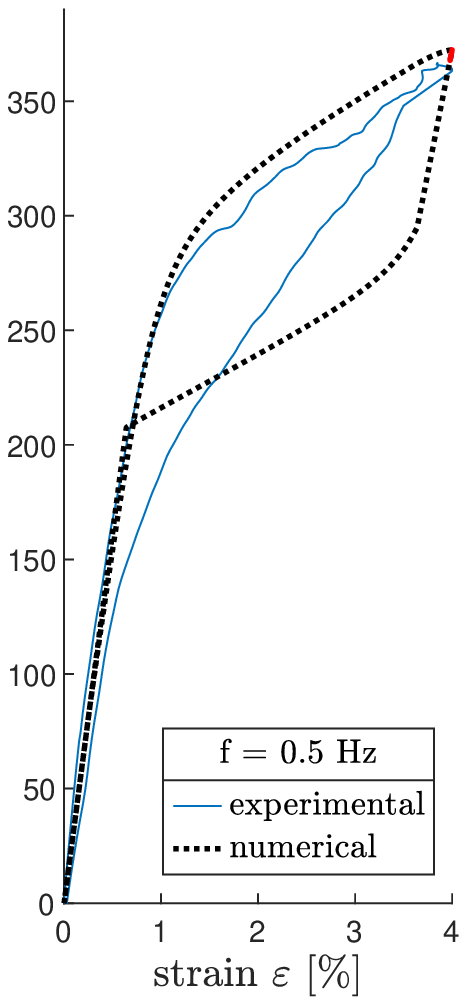}}
	\caption{Comparison of experimental data with numerical results including entropy-change effects on $\varepsilon_s^{MA}(\eta)$. Red-marked solid line: martensitic linear-elastic section. $\varepsilon=4\;\%$. $f=0.05 (a)/0.1 (b)/0.5 (c)$~Hz. $L=150$~mm. $d=0.2$~mm. $\sigma_0=134.9$~MPa. $T=22.5\;{^\circ}$C.}
	\label{e}
\end{figure}

For increasing strain-rates the difference between experimental and the numerical reverse transformation finish stress level increases continuously. Next, Fig.~\ref{e}, shows the included entropy-change effect on $\varepsilon_s^{MA}(\eta)$. To illustrate the effects of the entropy change on $\varepsilon_s^{MA}(\eta)$, we compare Fig.~\ref{e} with Fig.~\ref{No} and conclude that the linear-elastic section decreases for increasing strain-rate, when entropy-change effects are included in the numerical model. Nevertheless, the reverse transformation shape remains nearly the same, even though the phase transition is introduced at a higher stress level. The red-marked linear elastic section marks the calculated area between the beginning of unloading condition and the beginning of the reverse transformation calculation. Actually, significant changes in the reverse transformation arise firstly after activating both the entropy dependent speed parameter $\beta(\eta)$ and the reverse transformation finish strain $\varepsilon_f^{MA}(\eta)$. Accordingly, further entropy dependent adjustments are necessary for a more accurate calculation of martensitic phase stability.
 

\begin{figure}[] 
	\centering
	\subfloat[]{
		\includegraphics[scale=0.6]{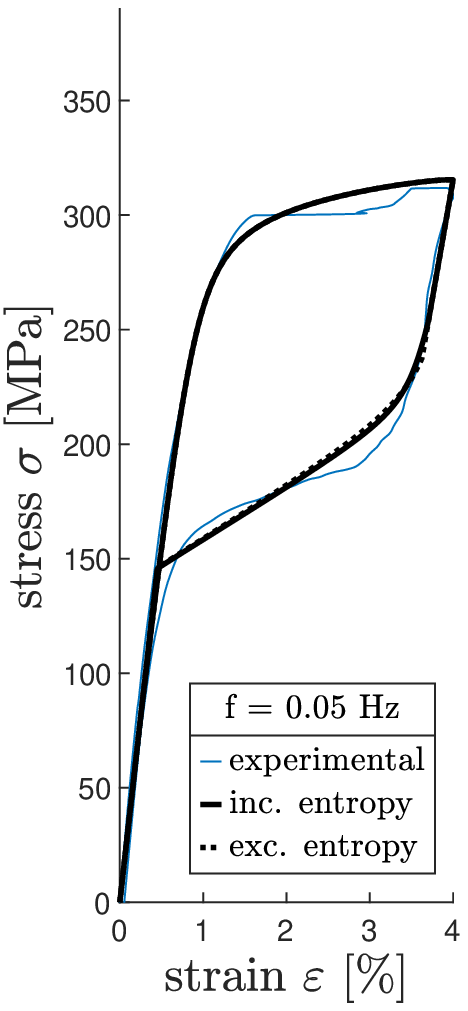}}
	\subfloat[]{%
		\includegraphics[scale=0.6]{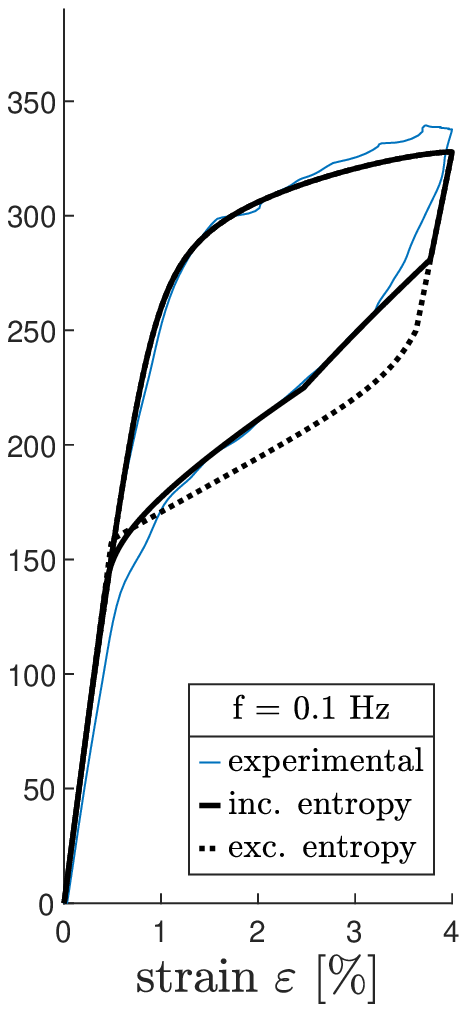}}
	\subfloat[]{%
		\includegraphics[scale=0.6]{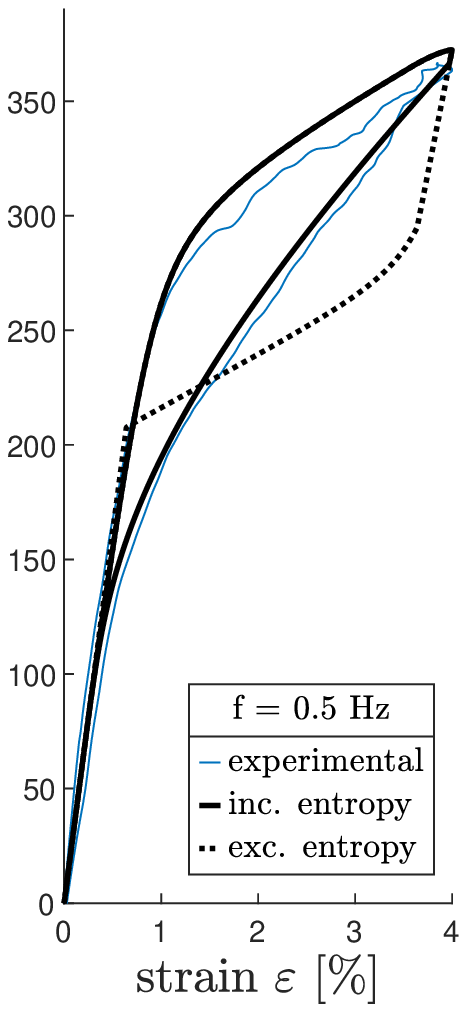}}
	\subfloat[]{%
		\includegraphics[scale=0.6]{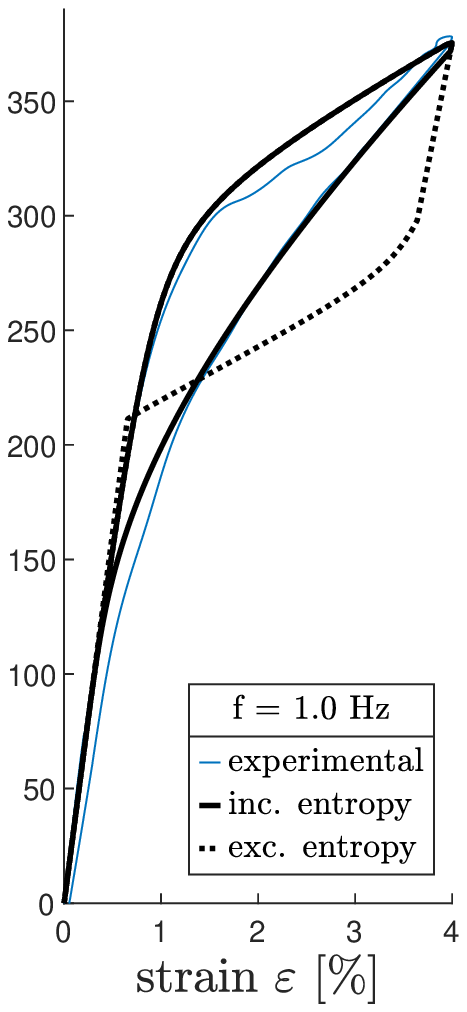}}
	\subfloat[]{%
		\includegraphics[scale=0.6]{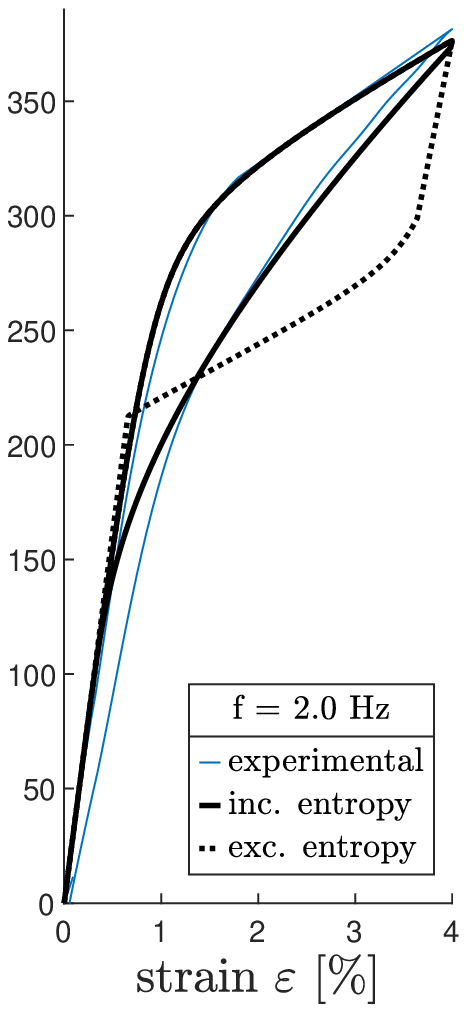}}
	\caption{Comparison of experimental data with numerical results both excluding and including entropy change. Thin solid line (blue): experimental investigation. Thick solid line (black): numerical results; including entropy change. Dotted line: numerical results; excluding entropy change. Entropy-change effects on $\varepsilon_{s,f}^{MA}(\eta)$ and $\beta^{MA}(\eta)$. $\varepsilon=4\;\%$. $f=0.05 (a)/0.1 (b)/0.5 (c)/1.0 (d)/2.0 (e)$~Hz. $L=150$~mm. $d=0.2$~mm. $\sigma_0=134.9$~MPa. $T=22.5\;{^\circ}$C.}
	\label{All}
\end{figure}

To incorporate the change in reverse transformation curve, we introduce the rewritten speed parameter $\beta^{MA}$ as a function of $\eta$. In fact, $\beta^{MA}(\eta)$ defines the shape of the cubic evolutionary equation for the reverse transformation and increases with increasing entropy, see Eq.~\ref{eq:11}. Fig.~\ref{All} shows the numerical results both including and excluding entropy changes, and compares the numerical to the experimental results. Here, $\varepsilon_{s,f}^{MA}(\eta)$ and $\beta^{MA}(\eta)$ are included in the calculation. The linear elastic area decreases for increasing entropy by way of updating $\varepsilon_{s}^{MA}(\eta)$. Simultaneously, the reverse transformation finish stress decreases for increasing entropy, since $\varepsilon_{f}^{MA}(\eta)$ is included in the calculation. Last, the speed parameter $\beta^{MA}(\eta)$ increases for increasing entropy and thus effectuates a reverse transformation slope change. Precisely, for quasi-static excitation in Fig.~\ref{All}(a) the numerical results with both excluded and included entropy change have a convex reverse transition slope and can represent the experimental results plausible. For a frequency of $0.1$  Hz the experimental results already reveal a decrease of the linear-elastic section and thus a steeper reverse transition slope, Fig.~\ref{All}(b). Comparing the numerical results of Fig.~\ref{All}(b), we see the importance of $\varepsilon_{s}^{MA}(\eta)$. Besides, the change from a convex to an almost straight reverse transformation curve, we can also observe an earlier initiation of the reverse transition. Indeed, the prior initiation arises from the entropy dependent calculation of $\varepsilon_{s}^{MA}(\eta)$, while the change of the slope comes from $\beta^{MA}(\eta)$. Even with the slight frequency increase from $0.05$ to $0.1$ Hz, the overestimation of the hysteresis surface can happen but can be reduced by including entropy changes in the calculation. In addition to the previously mentioned advantages of the included entropy changes, we now focus on the high strain-rate results in Fig.~\ref{All} (c), (d) and (e) and the influence of $\varepsilon_{f}^{MA}(\eta)$. All three results for high strain-rates are appropriate to illustrate the difference between $\varepsilon_{f}^{MA}$ and $\varepsilon_{f}^{MA}(\eta)$. Comparing the finish reverse transformation stress $\sigma_{f}^{MA}$ of both numerical results for a frequency of $0.5$ Hz, we observe a difference of more than $50$ MPa. In particular, by excluding the entropy change, the numerical model overestimates the experimental $\sigma_{f}^{MA}$ and, moreover, the phase transition to the linear-elastic austenitic part becomes abrupt. This abrupt transition does not fit to the experimental results. However, the numerical model with including entropy change effect calculates $\sigma_{f}^{MA}$ fairly well and thus also allows a smooth transition from the phase transition to the linear-elastic austenitic section. In conclusion, the inclusion of the rate-dependent entropy change for the reverse transition improves the calculation accuracy. Furthermore, the austenite transformation changes for higher strain-rate from a convex to a slightly concave curve due to the entropy dependent calculation of $\beta^{MA}(\eta)$. Moreover, the initial (finish) level for reverse transformation rises (decreases) for increasing strain-rates. Consequently, the advancements on the one-dimensional material model enable, especially for high strain-rates, a more accurate calculation of the hysteresis surface without negative impact on the numerical efficiency of the algorithm, which both will be proved numerically in the following.

\begin{figure}[h]
	\begin{center}
		\includegraphics[scale=0.78]{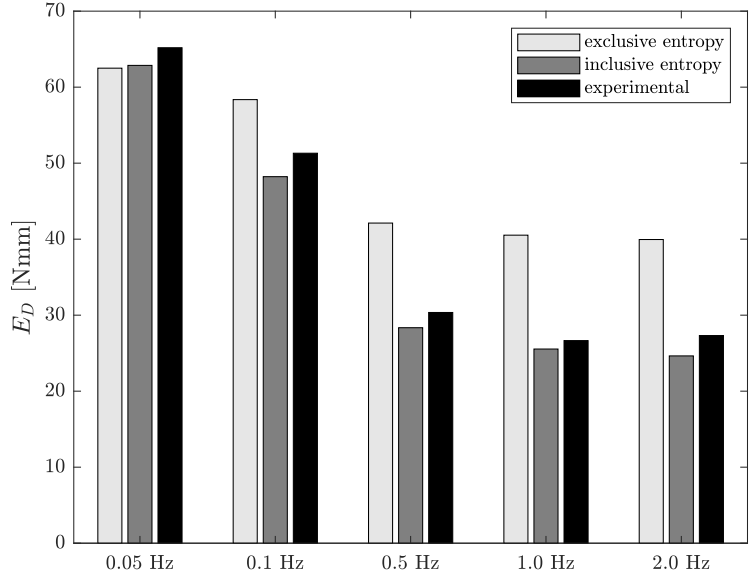}
		\caption{Dissipated energy per load cycle. Comparison of experimental data with numerical results both excluding and including entropy change. $\varepsilon=4\;\%$. $f=0.05/0.1/0.5/1.0/2.0$~Hz. $L=150$~mm. $d=0.2$~mm. $\sigma_0=134.9$~MPa. $T=22.5\;{^\circ}$C.}
		\label{w}
	\end{center}
\end{figure}

To illustrate the improvement of the numerical model, Fig.~\ref{w} visualizes the dissipated energy $E_D$ in [Nmm] of one load cycle for frequencies of 0.05, 0.1, 0.5, 1.0 and 2.0 Hz. For each frequency we compare the dissipated energy of the experimental data with numerical results with both excluded and included entropy changes in the calculation. The dissipated energy corresponds to the hysteresis surface calculated from the force-displacement curve. In general, Fig.~\ref{w} shows, for all previously in Fig.~\ref{All} shown three results, the decreasing energy dissipation of the superelastic SMA for an increasing strain-rate. Furthermore, these results reveal a significant difference of the dissipated energy between the quasi-static case with $f = 0.05$ Hz and the dynamic case with $f \ge 0.5$ Hz. The accuracy difference of both calculation does not change for the dynamic cases. Taking into account the quasi-static case, both numerical models slightly underestimate the energy dissipation of the experimental results. However, for increasing strain-rate the numerical results with excluded entropy-change effects always overestimate the energy dissipation compared to the experimental results. In fact, the difference continues to increase with increasing strain-rate. In contrast, the dissipated energy of the numerical calculation, with considered entropy effects, decreases analogously to the experimental results. Even though the entropy change included model slightly underestimates the energy potential compared to the experimental results, the results differ significantly less. Accordingly, the numerical model with including entropy change effects calculates the hysteresis surface more accurately than the comparative model and thus conforms with the experimental data.

\begin{figure}[] 
	\centering 
	\subfloat[]{\includegraphics[scale=0.5]{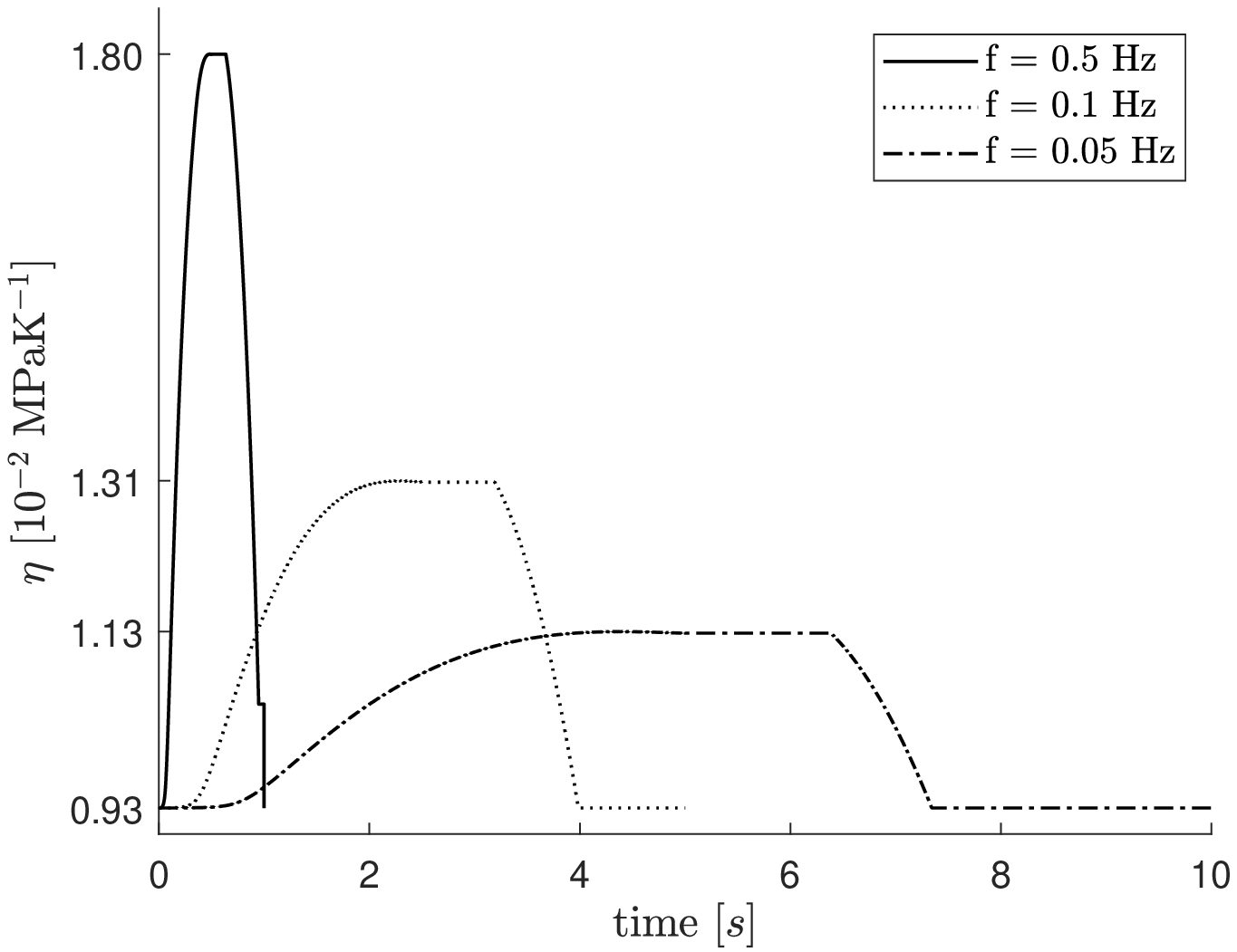}}\quad
	\subfloat[]{\includegraphics[scale=0.5]{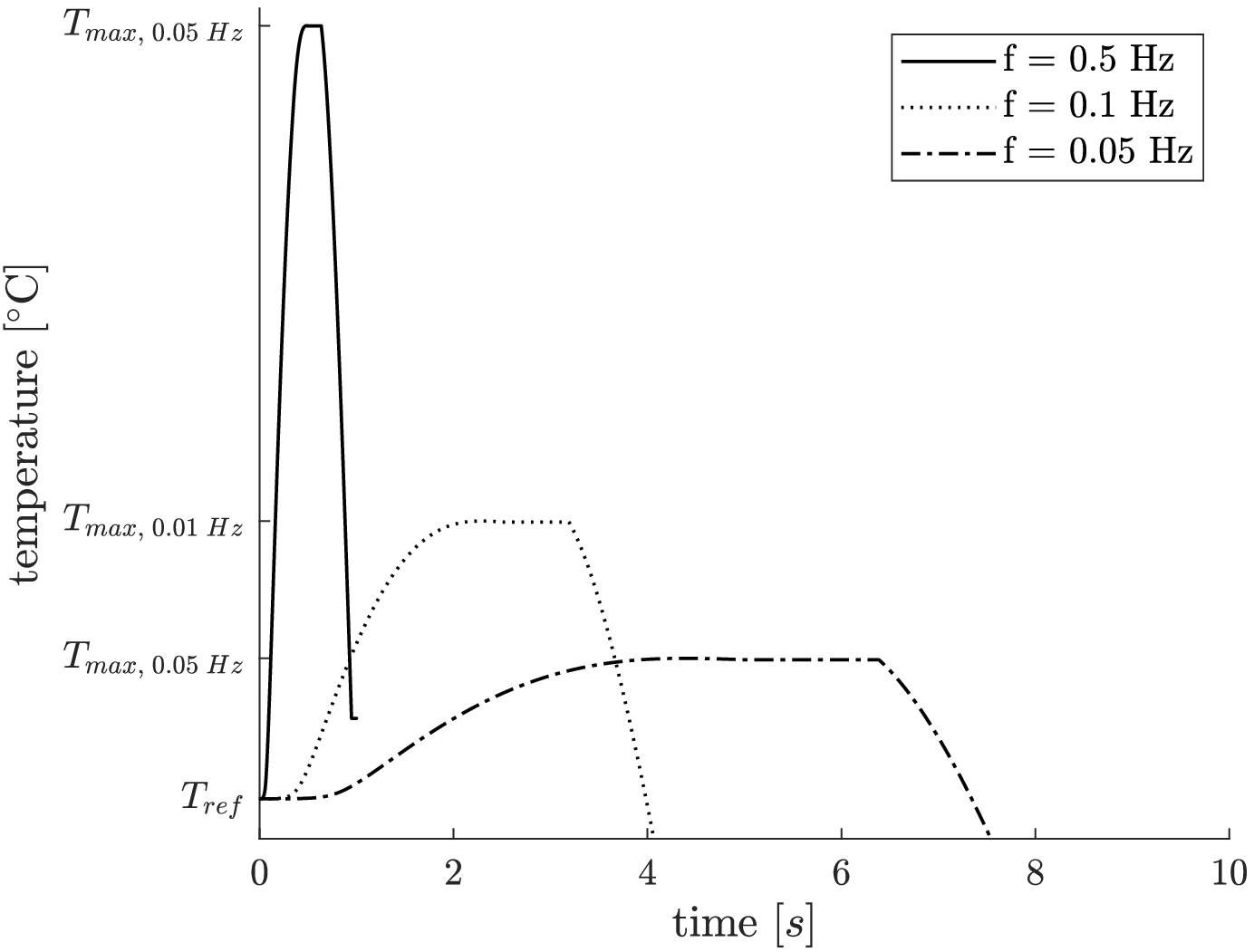}}\\
	\subfloat[]{\includegraphics[scale=0.5]{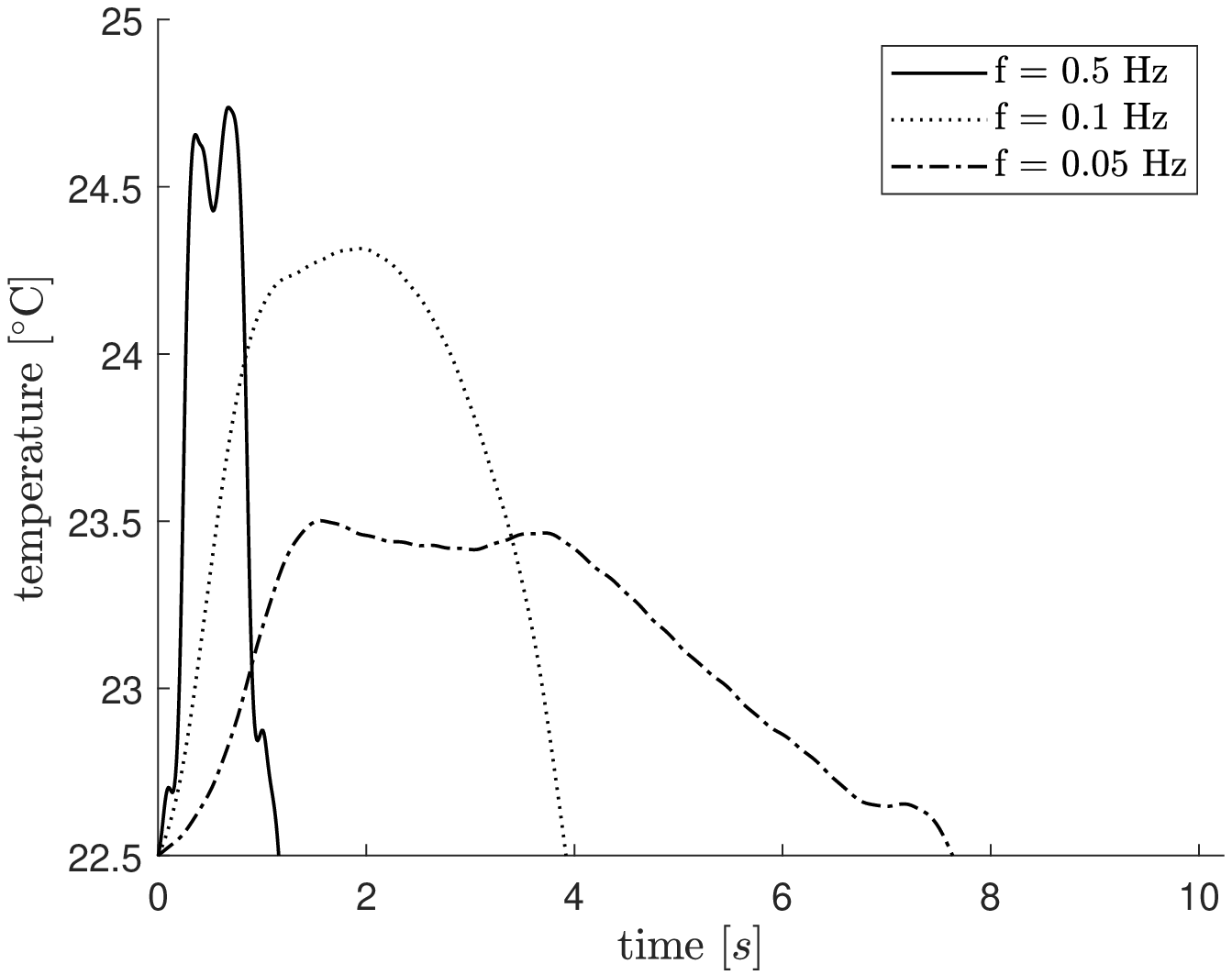}}\quad
	\subfloat[]{\includegraphics[scale=0.5]{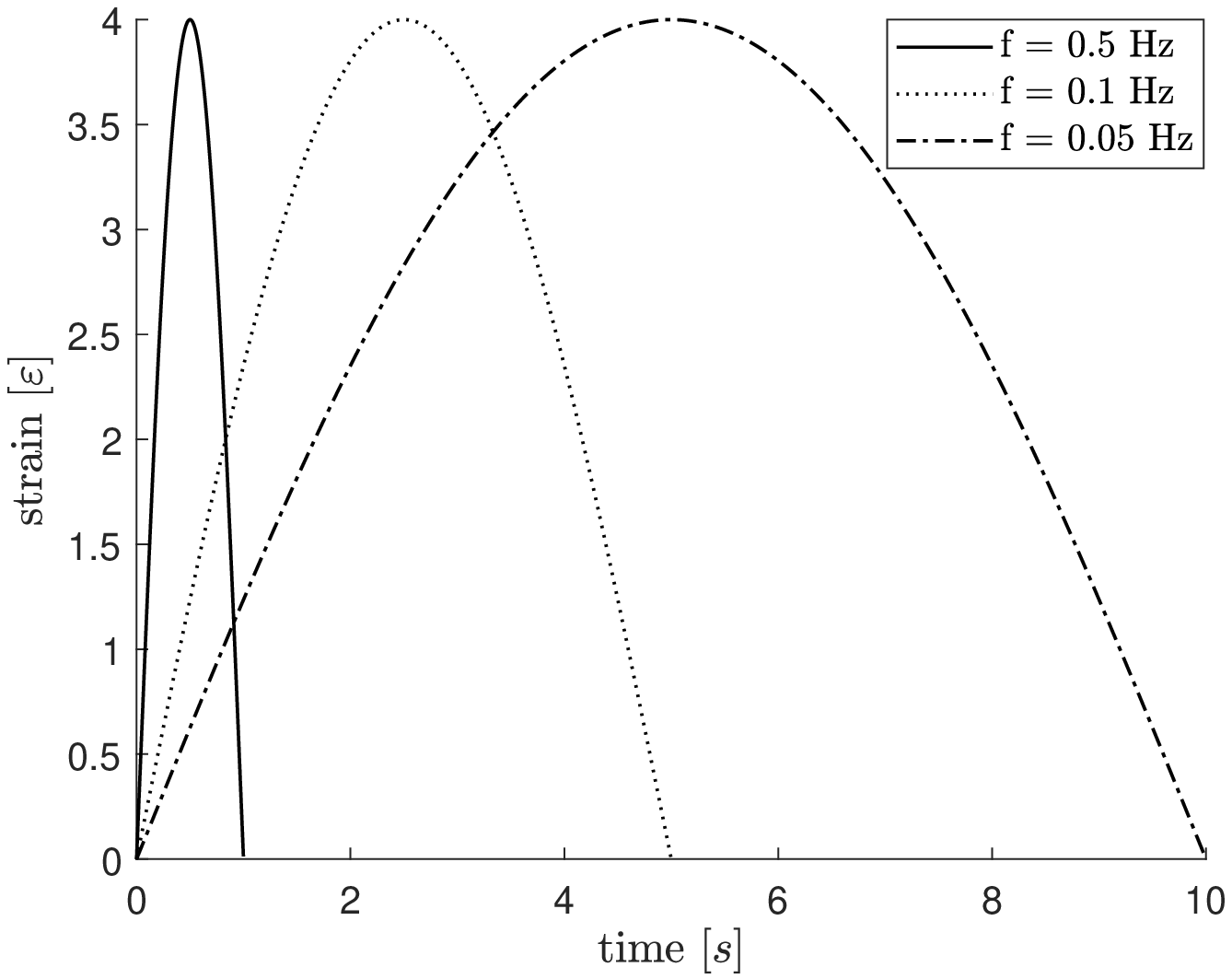}}
	\caption{Comparison of (a) numerical entropy calculation; (b) simulation temperature evolution; (c) experimental temperature evolution $T_{E}$; (d) strain-time history. $\varepsilon=4\;\%$. $f=0.05 (a)/0.1 (b)/0.5 (c)$~Hz. $L=150$~mm. $d=0.2$~mm. $\sigma_0=134.9$~MPa. $T=22.5\;{^\circ}$C.}
	\label{Te}
\end{figure}

Next, we concentrate on the calculation time of the numerical models. The calculation time using a standard PC differs between the two numerical models for one load cycle with 2~Hz in average 0.006~seconds, which is negligibly small.

After discussing the numerical and experimental results concerning the hysteresis surface and damping capacity, we will subsequently deal with entropy- and temperature evolution per se. For this purpose, the numerical entropy- and the simulation temperature evolutions are shown in Fig.~\ref{Te} (a) and (b). The experimental temperature results are illustrated in Fig.~\ref{Te} (c) and the related strain curves in Fig.~\ref{Te} (d). Each subdiagram illustrates the evolution for one load cycle with a frequency of 0.05, 0.1 and 0.5 Hz. The temperature and the entropy increases for increasing strain-rate. As mentioned prior, the calculated simulation temperature $T$ is not equal to the experimentally measured temperature $T_{E}$. For the entropy-change dependent calculation, the slope and the course of the simulation temperature are still decisive, therefore, Fig.~\ref{Te} (b) does not include any specific values. This diagram is intended to illustrate that the simulation temperature evolution corresponds approximately to the measured experimental temperature development and, moreover, the diagram has the purpose to demonstrate the influence of the temperature behavior on the numerical entropy evolution. Even though the entropy evolution (a) and the simulation temperature evolution (b) are quite similar, they differ explicitly in the start and finish values. Considering one completed load cycle, the entropy will always start and finish at $\eta_{A}$, because as soon as the material is completely in its austenite shape again, there is no increase in entropy due to the martensitic instability anymore. More in detail, at zero-stress level and an ambient temperature over the critical austenite finish temperature, the SMA is always in the superelastic state, where the austenitic grid-structure is energetically more efficient. In this state, the initial entropy condition $\eta_{A}$ always applies. A stress induced phase transformation leads to an increased entropy evolution during loading. The simulation temperature, on the other hand, will initially start at $T_{ref}$, but, depending on the excitation, the temperature at the end of one load cycle can be both higher or lower compared to the initial temperature. Thus, for the next load cycle the temperature calculation will start on the appropriate temperature level, while the entropy calculation will start at $\eta_{A}$ again. Last, the experimental temperature data (c) results from the T-type thermocouple installed 60~mm distanced from the right-side fixture. The graph is shifted to the room temperature of $T=22.5\;{^\circ}$C as initial temperature. Although, the wire is only 0.2 mm in diameter the temperature increases for one load cycle with $0.5$ Hz and a strain amplitude of $4\;\%$ more then $2\;{^\circ}$C. In summary, the temperature curve can be qualitatively calculated with the numerical model and further the comparison of the subdiagrams in Fig.~\ref{Te} verifies the plausibility of the simulation temperature and the entropy evolution.

\section{Conclusions}
In this paper, the influence of the rate-dependent phase-stability during the martensitic transformation of superelastic SMA-wires on the hysteresis surface is investigated both experimentally and considered numerically. To consider the rate-dependent phase-stability numerically, we advanced a one-dimensional macroscopic model by introducing an improved free energy formulation for the phase transition of superelastic SMA-wires. The proposed free energy formulation enables the rate-dependent entropy change calculation in the material to consider the decreasing martensitic phase-stability for increasing strain-rates. In brief, the entropy change calculation enables a strain-rate dependent calculation of the initial and finish strain level for the reverse transformation. The advancement further influences the reverse evolutionary equation and thus the shape of martensitic-austenitic transformation. Hence, the change from a convex to a concave reverse transition shape for increasing strain-rates can be calculated more accurately. The significant influence of the strain-rate on the reverse transformation is shown by experimental results. In particular, we performed cyclic-tensile tests with varying strain-rates on NiTi-wires, using a uniaxial horizontal shaking table. Furthermore, the fundamental findings of the previous research on the evolution of martensite transformation bands are gathered. Thus, both the relationship between strain-rate and martensitic phase stability and the necessity for a numerical consideration of this relationship have been noted. Furthermore, the experimental results are utilized to validate the improved one-dimensional model. Besides the comparison of the hysteresis surface of the numerical and the experimental stress-strain diagrams, we also compared the experimental and the simulation temperature evolution to determine the plausibility of the calculated entropy evolution. In conclusion, the rate-dependent thermomechanical modeling of the entropy changes in superelastic SMAs enables to consider the martensitic phase stability effects on the reverse transformation. This allows a more accurate calculation of the hysteresis surface, as shown in this paper by the presented calculations and experiments, with neither impairment of the numerical models simplicity nor the robustness of the solution algorithm.

The key features of the presented work are:
\begin{itemize}
	\item a free energy formulation incorporating the rate-dependent entropy change
	\item consideration of strain-rate dependent martensitic phase stability in an one-dimensional macroscopic model
	\item introduction of an entropy-dependent speed-parameter $\beta^{MA}(\eta)$
\end{itemize}
\section*{Acknowledgments}
The authors would like to express their sincere appreciation for the financial support from 
the German Research Foundation (Deutsche Forschungsgemeinschaft, DFG) with the 
grant number: KL 1345/12-1.

\end{document}